\documentclass[pre,twocolumn,preprintnumbers,amsmath,amssymb,floatfix,longbibliography,superscriptaddress]{revtex4-2}
\usepackage{graphicx}
\usepackage{epstopdf}
\usepackage{psfrag}
\usepackage{hyperref}
\usepackage{multirow}
\usepackage[sort&compress]{natbib}
\usepackage{amssymb}
\usepackage{amsmath}
\usepackage{amsthm}
\usepackage{bbold}
\usepackage{bbm}
\usepackage{enumerate}
\usepackage{textcomp}
\usepackage{verbatim}
\usepackage{soul}
\usepackage{ulem}
\usepackage[caption=false]{subfig}
\usepackage{orcidlink}
\usepackage{hyperref}       

\newcommand{\vare}{\varepsilon}

\newcommand{\beq}{\begin{equation}\begin{aligned}}
\newcommand{\eeq}{\end{aligned}\end{equation}}

\usepackage{xcolor}

\definecolor{AC}{rgb}{1, 0.2, 0.7}
\definecolor{SD}{rgb}{1,0,1}

\graphicspath{{./}{./figs/}}

\hypersetup{
    colorlinks,
    linkcolor=blue,
    citecolor=blue,
    urlcolor=blue,
    breaklinks=true 
}

\begin{document}

\title{Discrete holography and density of states in the crossover from hyperbolic to Euclidean lattices}

\author{Mireia Tolosa-Sime\'{o}n\,\orcidlink{0000-0003-4032-0132}}
\affiliation{Theoretische Physik III, Ruhr-Universit\"{a}t Bochum, D-44801 Bochum, Germany}

\author{Igor Boettcher\,\orcidlink{0000-0002-1634-4022}}
 \affiliation{Department of Physics, University of Alberta, Edmonton, Alberta T6G 2N8, Canada}
 \affiliation{Theoretical Physics Institute, University of Alberta, Edmonton, Alberta T6G 2N8, Canada}
 \affiliation{Quantum Horizons Alberta, University of Alberta, Edmonton, Alberta T6G 2N8, Canada}

\begin{abstract}
We study tight-binding models in the crossover from hyperbolic to Euclidean lattices, realized through the successive insertion of Euclidean defects into hyperbolic lattices.
We analyze how the holographic two-point boundary correlation function and bulk density of states evolve as defects are gradually introduced.
We find that bulk properties are strongly affected by the presence of Euclidean defects, whereas boundary observables remain remarkably robust even at high defect fractions.
This robustness indicates that essential features of boundary physics on hyperbolic lattices, which capture aspects of anti de-Sitter/conformal field theory (AdS/CFT)-like dualities in discrete systems, can be reproduced both experimentally and numerically without requiring perfectly hyperbolic lattices, thereby reducing the system size needed for implementation.
\end{abstract}

\maketitle

\section{Introduction}

Hyperbolic lattices recently spurred immense interest in condensed matter physics, quantum information, and high-energy physics due to their potential to emulate negatively curved spaces in the laboratory \cite{kollar2019hyperbolic,PhysRevX.10.011009,Boettcher2020,Lenggenhager2021,zhang2022observation,chen2023hyperbolic,zhang2023hyperbolic,huang2024hyperbolic,patino2024hyperbolic,fleury2024,Xu25}, experimentally realize holographic dualities \cite{PhysRevD.102.034511,brower2021lattice,Brower2022,10.21468/SciPostPhys.15.5.218,PhysRevLett.130.091604,chen2023adscft,Dey2024,ChenType2}, implement efficient quantum error correcting codes \cite{pastawski2015holographic,Breuckmann_2016,Breuckmann_2017,Lavasani_2019,jahn2021holographic,AliError,RayanError}, or host novel phases of matter \cite{Daniska:2016,Yu2020,Zhu:2021,maciejko2020hyperbolic,Bienias2022,Maciejko2022,Stegmaier2022,Liu2022,cheng2022band,PhysRevE.106.034114,PhysRevB.106.155146,PhysRevB.106.155120,kienzle2022,Urwyler2022,PhysRevB.107.125302,PhysRevB.107.165145,PhysRevB.107.184201,PhysRevB.108.035154,Chen2023Symmetry,PhysRevLett.131.226401,PhysRevB.109.L041109,nagy2023,tummuru2023hyperbolic,petermann2023eigenmodes,PhysRevLett.133.066101,PhysRevLett.133.146601,Li_2024,10.21468/SciPostPhys.17.5.124,PhysRevB.110.245117,PhysRevLett.134.076301,PhysRevB.111.L121108,Bitan1,PhysRevB.111.125125,s25y-s4fj,gf6q-kxh3,Bitan2,mx1t-74dm,3ylt-q7r3,Bitan3,rr6b-8536,Bitan4,hu2025generalization,YUAN20253146,SelaGoldstone,z2zk-m4c3}. 
In their simplest and most frequently studied form, hyperbolic lattices are $\{p,q\}$-tilings of the plane, made from regular $p$-gons with $q$ neighboring sites per vertex, with $(p-2)(q-2)>4$ \cite{kollar2019line,Boettcher2022,schrauth2023hypertiling,PhysRevLett.131.176603,PhysRevD.111.046001,thurn25hypertiling}. 
Remarkably, these tessellations have a crystallographic structure that is encoded in noncommutative translation groups \cite{BALAZS1986109,maciejko2020hyperbolic,Boettcher2022}. 
Classical or quantum particles moving on hyperbolic lattices experience an effective negative curvature. Euclidean $\{p,q\}$-tessellations satisfy $(p-2)(q-2)=4$, resulting in particles that effectively experience a flat or non-curved background. 
While there are infinitely many solutions $p,q$ for the hyperbolic case, the three Euclidean solutions, given by $\{3,6\}$, $\{4,4\}$, and $\{6,3\}$, correspond to the familiar triangular, square, and honeycomb lattices, respectively.

The mentioned applications of hyperbolic lattices mostly originate from the following property. 
If we cut out a finite flake (meaning a connected planar subgraph) from a hyperbolic tiling, then its number of sites grows exponentially in its graph diameter and a macroscopic fraction of sites resides on the boundary of the flake. 
This also implies that the shortest path to go from one boundary site to another typically passes through the bulk of the flake and that its graph-geodesic length grows like the logarithm of the distance along the boundary. 
In holographic toy-models defined on hyperbolic lattices, correlations akin to a conformal field theory (CFT) emerge on this macroscopically large boundary and the fact that geodesics go through the bulk implies a universal power-law scaling of the boundary-to-boundary correlator \cite{PhysRevD.102.034511,brower2021lattice,chen2023adscft,Dey2024}. 
In quantum error correction applications, these properties imply an efficient tradeoff between the number of vertices and edges in the code \cite{pastawski2015holographic}.

Keeping to the example of holographic CFTs emerging at the boundary of hyperbolic flakes, we may ask the following question: 
If the connectivity of hyperbolic lattices \textit{always} implies power-law scaling of boundary correlation functions, which in turn corresponds to the critical point of a second-order phase transition, how is it possible to go off criticality? 
After all, second-order phase transitions are rare in Euclidean space and generic many-body systems are non-critical. 
In contrast, on hyperbolic lattices, criticality appears to be generic.

\begin{figure*}[t!]
    \centering
    \includegraphics[width=\linewidth]{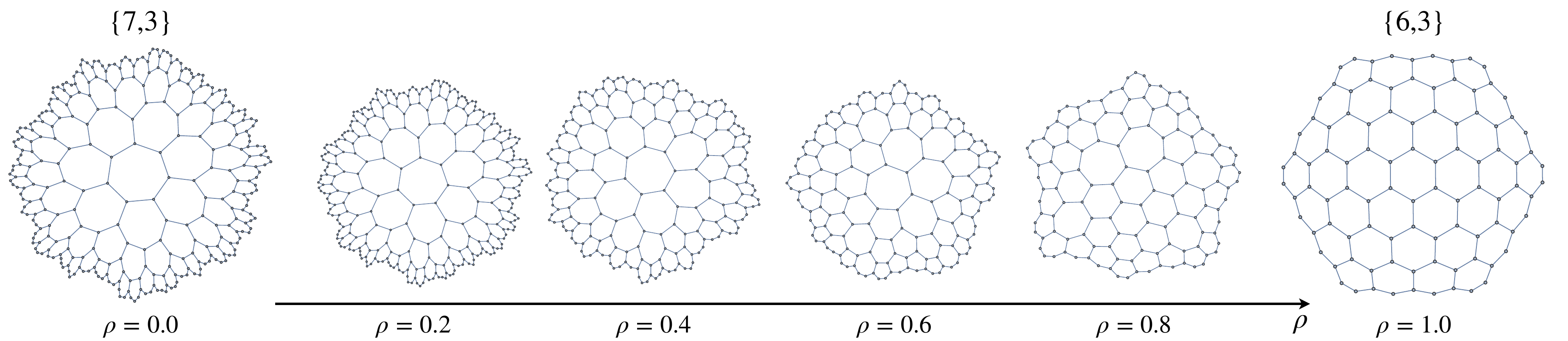}
    \caption{We show some realizations of the hyperbolic-to-Euclidean lattice crossover, where hexagonal faces are randomly inserted as defects into a hyperbolic $\{7,3\}$ graph, eventually producing a Euclidean $\{6,3\}$ honeycomb lattice. The fraction of hexagons is denoted by $\rho$. Note that the properties of tight-binding models on these graphs are independent of the physical coordinates of the vertices in the plane, which may thus be chosen arbitrarily. The connectivity of the graph, on the other hand, has crucial influence on the physical properties.}
    \label{fig:Crossover}
\end{figure*}

To address this question, we study in this work a family of planar graphs that interpolates between the hyperbolic and Euclidean limits. 
For concreteness, we analyze the crossover from hyperbolic $\{7,3\}$ or $\{8,3\}$ flakes to Euclidean $\{6,3\}$ honeycomb flakes by tuning a parameter $\rho\in[0,1]$ that represents the fraction of hexagons in the flake, as shown in Fig.~\ref{fig:Crossover}. 
The hexagons constitute defects in an otherwise pure hyperbolic lattice. 
We find that some properties of hyperbolic lattices disappear rapidly as $\rho >0$, such as sharp peaks in the density of states, while others are remarkably robust up to $\rho \lesssim 0.9$, like universal power-law behavior of boundary correlations. 
This, in turn, implies that applications of hyperbolic lattices do not require pure lattices, whose number of sites grows exponentially in the graph diameter, but defective lattices with fewer sites may be sufficient. 
This is important for any experimental realization, since creating more sites is typically more challenging. 

\section{Constructing graphs from tile-by-tile inflation}

Central to our analysis are planar graphs that are made from polygons whose number of sides can vary locally, while the coordination number, defined as the number of nearest neighbors, is 3 for each vertex. 
This should be viewed in contrast to regular $\{p,3\}$ tessellations, which are made entirely of $p$-gons. 
We construct these graphs as flakes of concentric rings of polygons where the number of sides of each polygon is drawn from a suitable random distribution.
Such graphs can be constructed through tile-by-tile inflation \cite{PhysRevX.10.011009,kollar2019line,Boettcher2020}, where the concentric rings are build one tile at a time and the value of $p$ for each such tile is specified individually. 
Knowing the sequence of $p$-values in the graph, from the central to the outer polygons in a counterclockwise fashion, allows us to construct the adjacency matrix. 
Here, for an undirected graph with $N$ sites labeled $\mu,\nu=1,\dots,N$, the adjacency matrix $A$ has components $A_{\mu\nu} = 1$ if $\mu$ and $\nu$ are neighboring sites (connected by an edge), and $A_{\mu\nu}=0$ otherwise.

We construct the graph as $l\geq 1$ concentric rings around a central polygon.
Therefore, the boundaries are smooth in the geometric sense and do not exhibit open polygons or dangling edges.
The $N$ sites of the graph are labeled by the index $\mu$ in a counterclockwise manner by starting on the first ring, then proceeding to the second ring, and so forth. 
Each site on the $l$th ring has two neighboring sites on the same ring, and a third neighboring site either on the inner $(l-1)$th  ring or the outer  $(l+1)$th ring. 
In the first case, we say that the site points ``inward'' or ``down'', while in the second case the site points ``outward'' or ``up''. 
Therefore, every site on the graph is either an up- or a down-site, and each ring is populated by a specific sequence of up- and down-sites. 
Constructing the adjacency matrix is then equivalent to building these up- and down-site sequences.

To create the adjacency matrices, we start from the first ring ($l=1$), which we assume to be a $p_{0}$-gon. This ring consists of $p_{0}$ up-sites.
We represent this sequence as a word
\begin{align}
 \label{tile1} \mathcal{W}_{l=1} = \{ \underbrace{\mbox{u},\dots,\mbox{u}}_{p_{0} \, \text{times}} \},
\end{align}
where ``$\mbox{u}$'' stands for up-site, and the ordering of sites is such that $\mu$ increases from left to right, i.e., $\{1,2,3,\dots,p_{0}\}$.
Given the word for the $l$th ring, the corresponding word for the $(l+1)$th ring is created using the following inflation rules: 
(i) For each site of the $l$th ring, a $p$-gonal tile is added as follows: Replace each entry ``$\mbox{u}$'' of the $l$th-ring word according to
\begin{align}
 \label{eq:uuRule}\mbox{u} &\mapsto \mbox{d},\underbrace{\mbox{u},\dots ,\mbox{u}}_{(p-4) \text{ times}}
\end{align}
if the next entry is ``$\mbox{u}$'', but replace ``$\mbox{u}$'' according to
\begin{align}
 \label{eq:udRule} \mbox{u} &\mapsto \mbox{d},\underbrace{\mbox{u},\dots ,\mbox{u}}_{(p-5) \text{ times}}
\end{align}
if the next entry is ``$\mbox{d}$''. 
For the last entry, we think of a cyclic word and compare with the first entry.
Importantly, the value of $p$ for every entry ``$\mbox{u}$'' can be chosen individually, giving the flexibility to construct any sequence of $p$-values as we go along the $l$th-ring word. 
(ii)~The entries ``$\mbox{d}$'' from the original $l$th ring word are discarded from the $(l+1)$th ring word. (iii) Finally, the adjacency matrix results from gluing together the words such that the first up-site of the $l$th word is connected to the first down-site of the $(l+1)$th word. The combination of steps (i)-(iii) amounts to adding $p$-gonal tiles one after the other at the up-sites, hence the name tile-by-tile inflation.
The up-down sequences can be used to determine the number of sites on each ring of a $\{p,3\}$ lattice as explained in Appendix~\ref{app:Recurrence}~\cite{Gu2011}.

\begin{figure*}[t!]
    \centering
    \includegraphics[width=0.85\linewidth]{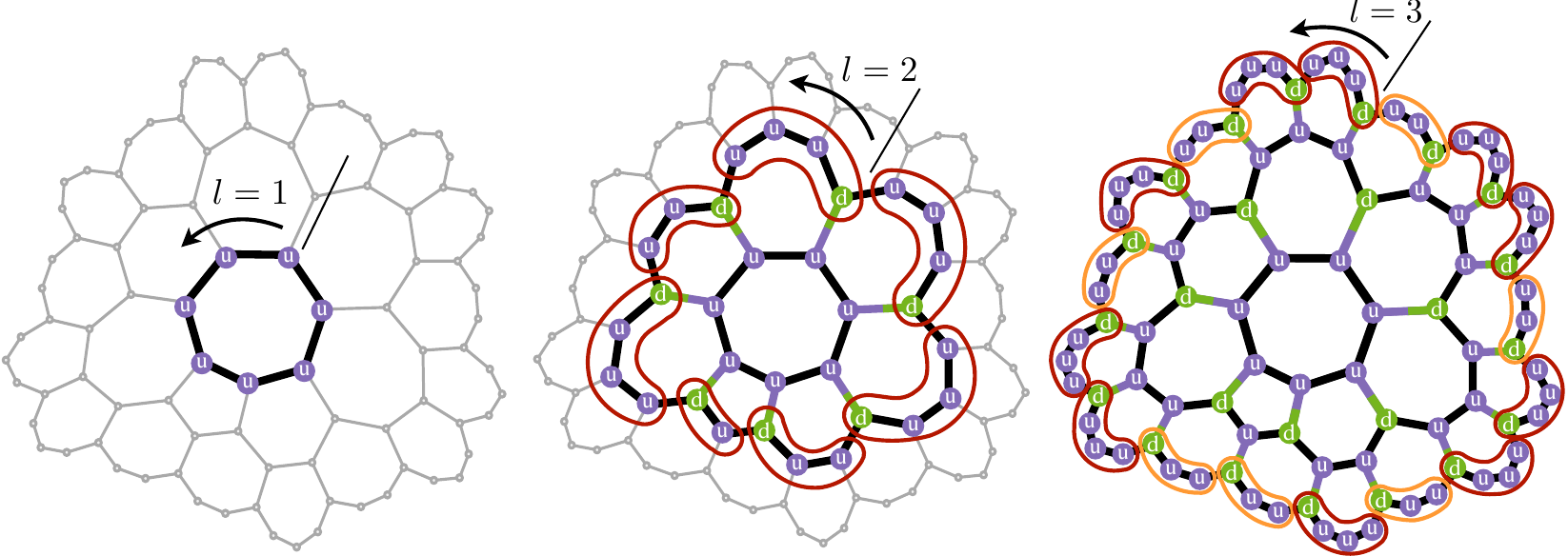}
    \caption{Construction of a defective $\{7,3\}$-flake with three concentric rings through tile-by-tile inflation. The graph features three defective tiles, one pentagon and two hexagons on the second ring, as given in the example in Eq.~\eqref{eq:ConstructionGraph}.
    The purple vertices labelled ``$\mbox{u}$'' correspond to sites pointing ``outward'' or ``up'', while the green vertices labelled ``$\mbox{d}$" correspond to sites pointing ``inward'' or ``down''.
    The arrow indicates that the rings are created counterclockwise by starting on the first ring, then proceeding to the second ring, and finally to the third ring.
    The dark red and the orange envelopes indicate that each entry ``$\mbox{u}$'' of the $(l-1)$th ring has been replaced according to Eq.~\eqref{eq:uuRule} and Eq.~\eqref{eq:udRule}, respectively, to create the $l$th ring}.
    \label{fig:ConstructionMatrixDefects}
\end{figure*}

As an example, consider a flake of three concentric rings comprised of a central 7-gon ($l=1$), surrounded by a ring that consists of (7, 6, 7, 5, 6, 7, 7)-gons ($l=2$), and a third ring consisting of 7-gons ($l=3$). The up-down words for the three rings are
\begin{align}
  \mathcal{W}_{l=1} = \{ {}&\underbrace{\mbox{u},\mbox{u},\mbox{u},\mbox{u},\mbox{u},\mbox{u},\mbox{u}}_7\},\\
   \mathcal{W}_{l=2} = \{{}&\mbox{d}, \underbrace{\mbox{u}, \mbox{u}, \mbox{u}}_{7-4}, \mbox{d}, \underbrace{\mbox{u}, \mbox{u}}_{6-4}, \mbox{d}, \underbrace{\mbox{u}, \mbox{u}, \mbox{u}}_{7-4}, \mbox{d}, \underbrace{\mbox{u}}_{5-4}, \mbox{d}, \underbrace{\mbox{u}, \mbox{u}}_{6-4}, \nonumber \\
 &  \mbox{d}, \underbrace{\mbox{u}, \mbox{u}, \mbox{u}}_{7-4}, \mbox{d}, \underbrace{\mbox{u}, \mbox{u}, \mbox{u}}_{7-4}\},
 \end{align}
\begin{align}
  \mathcal{W}_{l=3} = \{{}&\mbox{d}, \underbrace{\mbox{u}, \mbox{u}, \mbox{u}}_{7-4}, \mbox{d}, \underbrace{\mbox{u}, \mbox{u}, \mbox{u}}_{7-4}, \mbox{d}, \underbrace{\mbox{u}, \mbox{u}}_{7-5}, \mbox{d}, \underbrace{\mbox{u}, \mbox{u}, \mbox{u}}_{7-4}, \mbox{d}, \underbrace{\mbox{u}, \mbox{u}}_{7-5}, \nonumber\\
  & \mbox{d}, \underbrace{\mbox{u}, \mbox{u}, \mbox{u}}_{7-4}, \mbox{d}, \underbrace{\mbox{u}, \mbox{u}, \mbox{u}}_{7-4}, \mbox{d}, \underbrace{\mbox{u}, \mbox{u} }_{7-5}, \mbox{d}, \underbrace{\mbox{u}, \mbox{u}}_{7-5}, \mbox{d}, \underbrace{\mbox{u}, \mbox{u}, \mbox{u}}_{7-4}, \nonumber \\
  & \mbox{d}, \underbrace{\mbox{u}, \mbox{u}}_{7-5}, \mbox{d}, \underbrace{\mbox{u}, \mbox{u}, \mbox{u}}_{7-4}, \mbox{d}, \underbrace{\mbox{u}, \mbox{u}, \mbox{u}}_{7-4}, \mbox{d}, \underbrace{\mbox{u}, \mbox{u}}_{7-5}, \mbox{d}, \underbrace{\mbox{u}, \mbox{u}, \mbox{u}}_{7-4},\nonumber\\
  & \mbox{d}, \underbrace{\mbox{u}, \mbox{u}, \mbox{u}}_{7-4}, \mbox{d}, \underbrace{\mbox{u}, \mbox{u}}_{7-5}\}.
  \label{eq:ConstructionGraph}
  \end{align}
This example is a $\{7,3\}$-flake with three defective tiles, namely one pentagon and two hexagons on the second ring. 
The corresponding construction of the graph is shown in Fig.~\ref{fig:ConstructionMatrixDefects}.

While the tile-by-tile inflation allows to construct graphs where the values of $p$ vary arbitrarily along the sites of the rings, we use the method in this work to study the crossover from a hyperbolic $\{7,3\}$ or $\{8,3\}$ graph to a Euclidean $\{6,3\}$ graph by introducing a fraction of $\rho\in[0,1]$ hexagonal tiles. To explain the construction of the crossover graphs, we focus on $\{7,3\}\to \{6,3\}$ for concreteness, as depicted in Fig.~\ref{fig:Crossover}.

For a given value of $\rho\in[0,1]$, we start from a central $p$-gon, where $p$ is chosen with probability $\rho$ as $p=6$ and otherwise as $p=7$ (with probability $1-\rho$). This constitutes the first ring. We then add the second ring through tile-by-tile inflation, where the $p$ at each step (i) is chosen again with probability $\rho$ to be $p=6$, and $p=7$ otherwise. We continue this procedure until we arrive at a flake with a certain number $l_{\rm max}$ of rings, which is a graph $\mathcal{G}^{(k)}$ with adjacency matrix $A^{(k)}$ that we call a realization.
Since the $p$ are chosen randomly, the graph $\mathcal{G}^{(k)}$ has a number of sites $N_k=|\mathcal{G}^{(k)}|$ that is not known a priori. Furthermore, the fraction of hexagons $\hat{\rho}_k$ of the realization will not be exactly $\rho$. However, we can post-select those realizations where $\hat{\rho}_k$ is close to $\rho$ (for instance, in the interval $\rho\pm \epsilon$ with tolerance $\epsilon$). For each $\rho$, we create $M$ such realizations $\mathcal{G}^{(k)}$, $k=1,\dots,M$, and determine the mean value from
\begin{align}
 \bar{\rho} &= \frac{1}{M}\sum_{k=1}^M \hat{\rho}_k,
\end{align}
together with the standard deviation $\Delta \rho$. In practice, we choose $M=30$ realizations and a tolerance of $\epsilon=0.01$ for each $\rho$.

For any realization $\mathcal{G}^{(k)}$, the fraction $\hat{\rho}_k$ can be obtained by dividing the number of times $p=6$ was drawn by the total number of faces in the graph. Alternatively, for a graph with only hexagons and heptagons, $\hat{\rho}_k$ is the number of 6-cycles divided by the number of 6- and 7-cycles. Here, counting the cycles of the graph is a mathematical way of counting the faces of the tessellation.

\section{Hyperbolic-to-Euclidean crossover}

To study the hyperbolic-to-Euclidean crossover, we consider in this work a system of particles whose dynamics are described by a tight-binding Hamiltonian or nearest-neighbor hopping action.
The changes in behavior of this system as the underlying graph is changed reveals the crossover in a physical setting. 
Of course, an alternative would be to study the crossover in an entirely graph-theoretic framework by analyzing, for instance, the eigenvalues of the adjacency matrix or number and lengths of certain graph geodesics. 
Our physical model does, in fact, encode closely related information through the density of states (eigenvalues) and holographic boundary correlations (lengths of geodesics).

\subsection{Model}

We consider a real scalar field $\phi_\mu$, where the index $\mu=1,\dots,N$ labels the site index on the graph or flake~$\mathcal{G}$, and thus acts as the position variable. The dynamics of the system is described by the classical action
\begin{align}
\label{eq:Action} S &= -t\sum_{\langle \mu,\nu\rangle} \phi_\mu\phi_\nu + \sum_\mu \frac{\hat{m}^2}{2} \phi_\mu^2+S_{\rm int}\\
 \label{mod1} &= -\frac{t}{2} \sum_{\mu,\nu}  \phi_\mu A_{\mu\nu} \phi_\nu + \sum_\mu \frac{\hat{m}^2}{2} \phi_\mu^2+S_{\rm int},
\end{align}
where the parameter $t>0$ describes hopping between nearest-neighbor sites and $\hat{m}^2$ is a mass-like term. We give two parameterizations of the nearest-neighbor kinetic term, one using the common $\langle \mu,\nu\rangle$-notation, while the other utilizes the adjacency matrix with the factor of $1/2$ to remove double counting.
Interactions can be incorporated through $S_{\rm int}$ by adding local terms that are of cubic or higher power in the field $\phi$, for instance,
\begin{align}
 \label{mod2} S_{\rm int}  = \frac{u}{3!}\sum_\mu \phi_\mu^3 + \dots \,.
\end{align}
In the non-interacting limit, $S_{\rm int}=0$, the action $S$ describes the tight-binding dynamics of single particles on the graph $\mathcal{G}$. Since this action is quadratic in the fields, we can write $S= \frac{1}{2}\sum_{\mu,\nu}\phi_\mu H_{\mu\nu} \phi_\nu$ with the single-particle Hamiltonian matrix 
\begin{align}
 H = (H_{\mu\nu}) = -t A +\hat{m}^2\mathbb{1}.
\end{align}
The single-particle DOS and two-point propagators can be obtained from this matrix alone. The formula relating $\hat{m}^2$ and $m^2$ is given in Eq. (\ref{masseq1}). In the remainder of the paper, we set 
\begin{align}
t=1.
\end{align}

The model in Eq.~(\ref{mod1}) constitutes a universal low-energy action for many physical systems in statistical mechanics. 
By promoting the fields to creation and annihilation operators, quantum effects can be incorporated to model typical condensed-matter or high-energy physics systems. 
Crucially, the action (\ref{mod1}), when defined on a regular hyperbolic lattice, represents a toy-model for the holographic duality or AdS/CFT correspondence \cite{Dey2024}.
In this interpretation, the nearest-neighbor hopping term approximates the Laplace--Beltrami operator, $\square$, on the hyperbolic Poincaré disk with curvature radius $\ell$.
Concretely, we replace $\sum_\nu A_{\mu\nu} \phi_\nu \to (q+qh^2\ell^2\square)\phi_\mu$ in Eq.~\eqref{mod1}, yielding
\begin{equation}
    S= \frac{1}{2} \sum_{\mu}  \phi_\mu (-qh^2\ell^2\square -q + \hat{m}^2)\phi_\mu +S_{\rm int}.
\end{equation}
This relation follows from the discretization of the covariant Klein--Gordon operator on a negatively curved background and ensures a well-defined continuum limit~\cite{Boettcher2020,petermann2023eigenmodes}.
Choosing the mass-like parameter as 
\begin{align}
 \label{masseq1} \hat{m}^2 = q+qh^2\ell^2m^2,
\end{align}
where $h$ is the dimensionless lattice constant, the action reduces to
\begin{equation}
   \label{masseq2} S= \frac{1}{2} \sum_{\mu}  \phi_\mu (-\square + m^2)\phi_\mu +S_{\rm int}.
\end{equation}
Note that the value of $h^2\ell^2$ is fixed by the integers $p$ and $q$ \cite{kollar2019hyperbolic,Boettcher2020}.

The physical mass $m$ determines the scaling dimension of the two-point boundary operators.
Specifically, in the continuum AdS$_2$, the corresponding scaling dimension $\Delta$ satisfies $m^2 \ell^2 = \Delta(\Delta-1)$~\cite{witten1998,Maldacena1999}, whereas discretization effects lead to $h$-dependent corrections on hyperbolic lattices \cite{Dey2024}.
The interaction term $S_{\rm int}$ gives rise to nonvanishing three-, four-, and higher-order boundary correlation functions in the dual CFT \cite{Dey2024}. 
For instance, starting from Eq.~\eqref{mod2}, the three- and four-point functions would be $\propto u$ and $\propto u^2$, respectively. In contrast, if $S_{\rm int}=0$, then only the boundary two-point correlation function is nontrivial. 
Since higher-point correlation functions are not important for our study of the hyperbolic-to-Euclidean crossover, we work in the following with $ \label{mod3} S_{\rm int} =0$.
The underlying single-particle physics described by the quadratic part of the action is still very rich. 


\subsection{Holographic boundary correlations}
\label{sec:Holography}

\begin{figure}[t!]
    \centering
    \includegraphics[width=\linewidth]{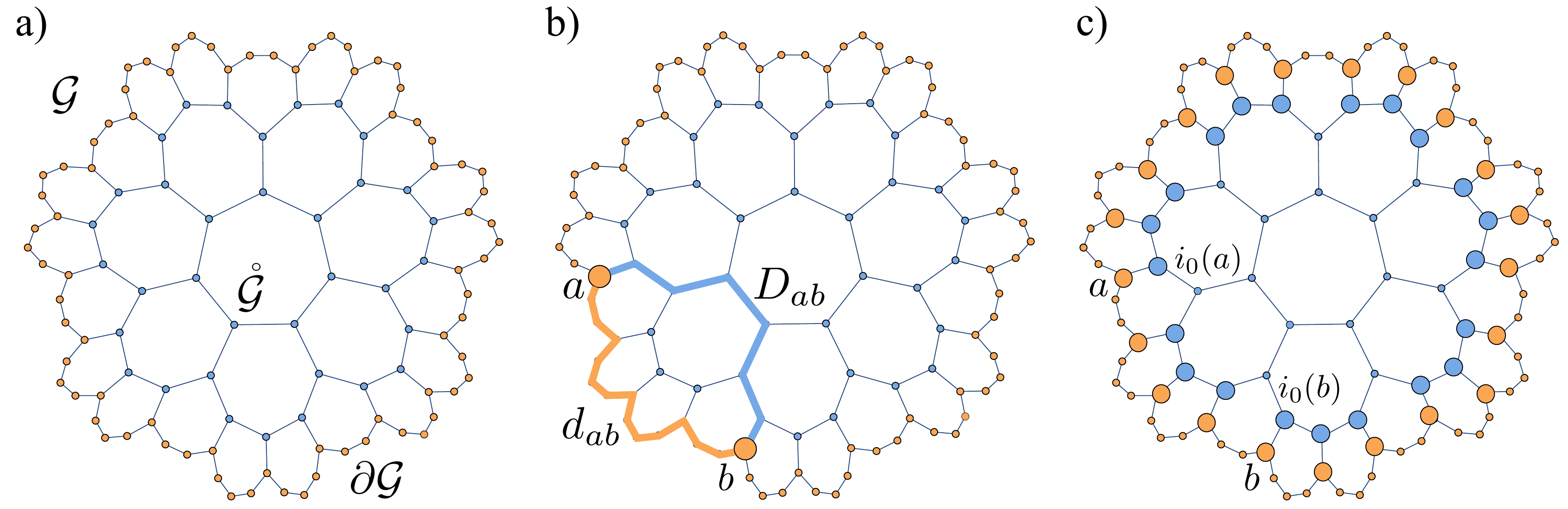}
    \caption{\textbf{a)}~Graph $\mathcal{G}$ divided into bulk, $\mathring{\mathcal{G}}$, depicted by blue vertices and boundary, $\partial \mathcal{G}$, depicted by orange vertices.
    \textbf{b)}~The graph distance $D_{ab}$ between two points $a$ and $b$ on the boundary  is indicated by a blue line, while the boundary distance $d_{ab}$ between the same two points is indicated by an orange line.
    \textbf{c)}~The large orange vertices constitute boundary sites $a$ with coordination number $q=3$, i.e., the down-sites on the boundary, while the large blue vertices constitute the corresponding bulk sites $i_0(a)$ that connect to them, i.e., the up-sites of the last interior ring.}
    \label{fig:BulkBoundary}
\end{figure}

We consider a flake with $N$ sites, represented by the graph $\mathcal{G}$ with adjacency matrix $A$. 
The sites on $\mathcal{G}$ are labeled by Greek letters $\mu,\nu\in\mathcal{G}$. 
We divide the graph into a bulk, $\mathring{\mathcal{G}}$, and a boundary, $\partial \mathcal{G}$, such that 
\begin{align}
\mathcal{G} = \mathring{\mathcal{G}} \cup \partial\mathcal{G}.
\end{align}
Bulk sites, labeled by Latin letters $i,j,\in\mathring{\mathcal{G}}$, are shown in Fig.~\ref{fig:BulkBoundary}a) as blue vertices, while boundary sites, labeled by $a,b\in\partial \mathcal{G}$, are depicted as orange vertices.
The bulk of a flake with $l$ concentric rings consists of the sites in the inner $(l-1)$ rings, whereas the boundary is identified with the outermost $l$th ring.

To understand the origin of holographic boundary correlations, it is important to distinguish the two natural distance measures on the whole graph $\mathcal{G}$ and on the boundary $\partial \mathcal{G}$.
On the one hand, we define the graph distance $D_{\mu\nu}$ between two sites $\mu,\nu$ by the minimal number of edges connecting $\mu$ and $\nu$ on the graph $\mathcal{G}$, also called graph-geodesic distance.
On the other hand, we define the boundary distance $d_{ab}$ between two boundary sites $a,b \in \partial \mathcal{G}$  as the minimal number of edges connecting $a$ and $b$ on the boundary $\partial\mathcal{G}$. Since $\partial\mathcal{G}$ consists of only one ring, we have
\begin{align}
 \label{holo2} d_{ab} = \text{min}\Bigl\{ |a-b|, \Bigl||\partial\mathcal{G}|-|a-b|\Bigr|\Bigr\}.
\end{align}
This corresponds to the shortest path along the outer ring, and thus is analogous to the angular separation between coordinates in the plane.

For two boundary sites $a,b\in\partial \mathcal{G}$, both distance measures $D_{ab}$ and $d_{ab}$ are well-defined. 
Clearly $D_{ab}\leq d_{ab}$, as exemplified in Fig.~\ref{fig:BulkBoundary}b).
Crucially, while $D_{ab}$ and $d_{ab}$ satisfy an approximately linear relation on Euclidean flakes, 
\begin{align}
 \label{holo4} D_{ab}\sim d_{ab}\ \text{(Euclidean)},
\end{align}
the relation is logarithmic on hyperbolic flakes, 
\begin{align}
 \label{holo5} D_{ab} \sim \log d_{ab}\ \text{(hyperbolic)}.
\end{align}
Eq.~\eqref{holo5} reflects the characteristic ``compression'' of distances in hyperbolic space, i.e., two boundary sites that are widely separated along the boundary may nevertheless be connected by only a few bulk steps.
This property underlies the emergence of power-law correlations on the boundary $\partial \mathcal{G}$.

\begin{figure}[t!]
    \centering
    \includegraphics[width=\linewidth]{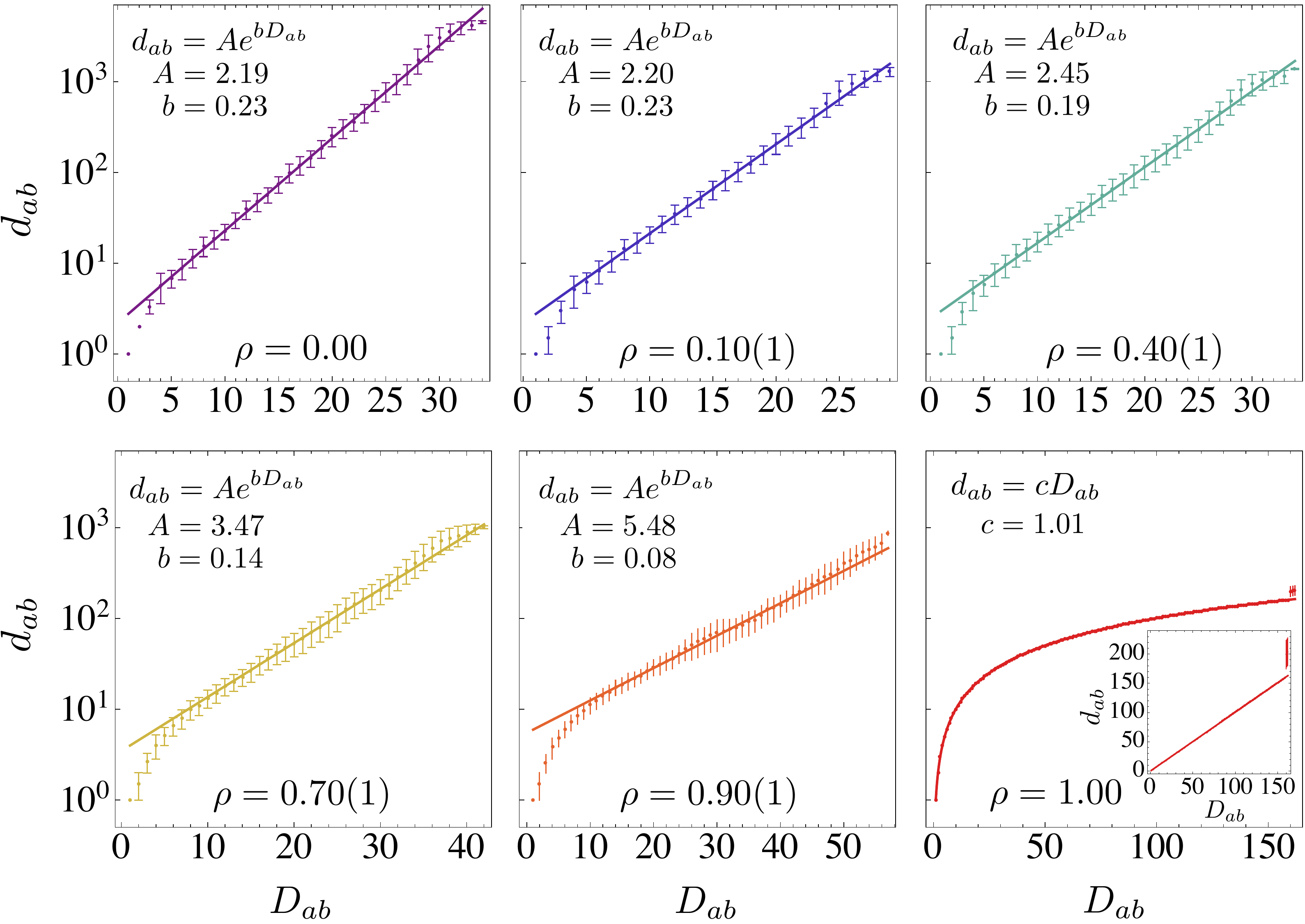}
    \caption{Boundary distance $d_{ab}$ as a function of the graph distance $D_{ab}$  in the hyperbolic-to-Euclidean lattice crossover ${\{7,3\}\to \{6,3\}}$, shown for several values of $\rho$. 
    For $\rho<1$ we apply the  fit from Eq.~\eqref{holo5}, while for $\rho =1.00$, we apply the fit from Eq.~\eqref{holo4}.
    The inset for $\rho =1.00$ displays the data on a linear scale, highlighting the linear Euclidean scaling behavior.
    The error bars indicate the standard deviation of the mean boundary distance computed for each graph distance. 
    The deviations result from the discreteness of the flakes and their rugged boundaries.
    The specific colors used for various $\rho$ in this plot are also employed in the remaining figures of the work.} 
    \label{fig:Distances}
\end{figure}

Fig.~\ref{fig:Distances} illustrates the relation between the different distance measures in Eqs.~\eqref{holo4} and \eqref{holo5} in the hyperbolic-to-Euclidean lattice crossover from ${\{7,3\}\to\{6,3\}}$.
It shows that the logarithmic relation between $D_{ab}$ and $d_{ab}$, characteristic of hyperbolic lattices, remains valid even in the presence of a large fraction of hexagonal (Euclidean) defects.
Therefore, a moderate number of Euclidean defects does not strongly affect the distance scaling, and thus the holographic boundary-to-boundary correlations, as we show below.
The relations between distance measures for the pure flakes with $\rho=0$ and $\rho=1$ are computed using $N\sim 10^4$ sites. 
Within the crossover regime and for each expected value of $\rho:=\langle \hat{\rho}\rangle$, we realize 30 flakes of random placements of hexagons, each with $N\sim 5000$ sites and a random $\hat{\rho}$, such that the variance is $\Delta \hat{\rho}=0.01$. 
The specific flake parameters are listed in Table \ref{TabDOS83}.

\begin{figure*}[t!]
    \centering
    \includegraphics[width=\linewidth]{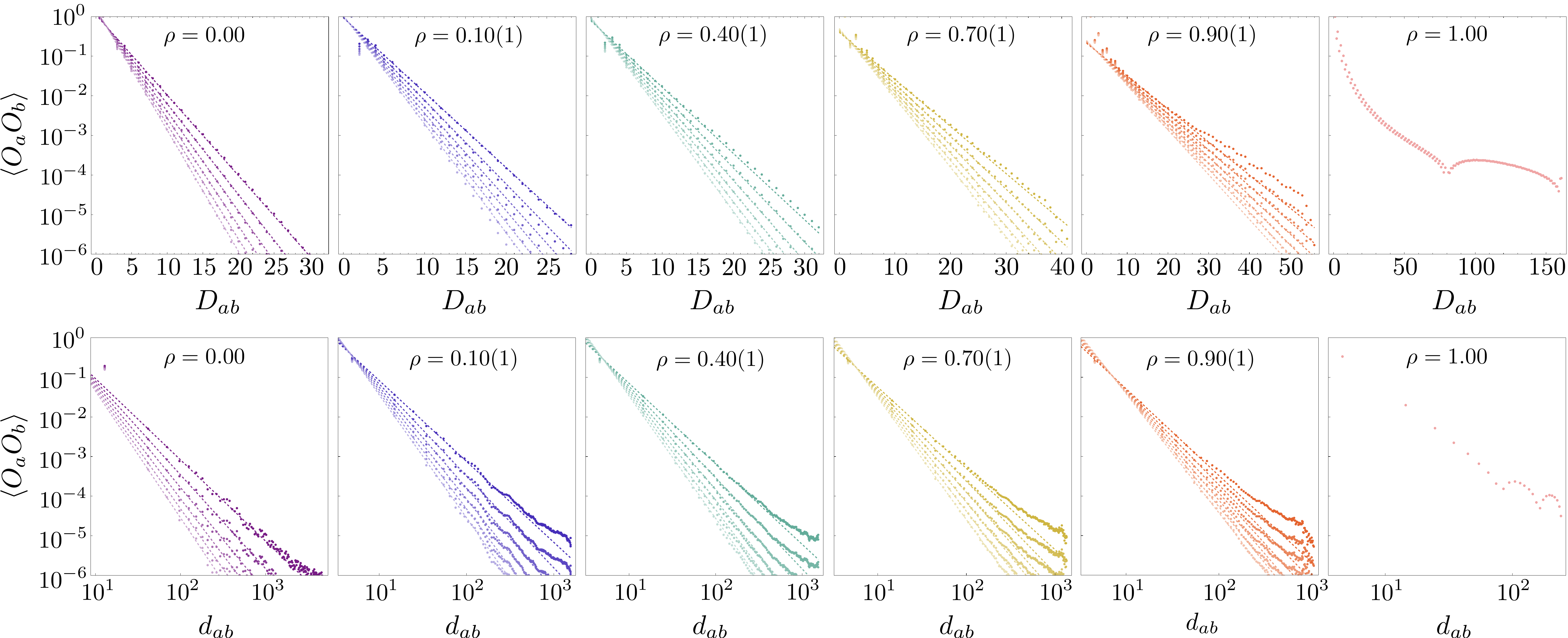}
    \caption{Boundary two-point correlation functions $\langle\mathcal{O}_a \mathcal{O}_b \rangle$ in the hyperbolic-to-Euclidean lattice crossover $\{7,3\}\to \{6,3\}$ as a function of graph distance $D_{ab}$ (upper row) and  boundary distance $d_{ab}$ (lower row), shown for several values of $\rho$. 
    Note that the upper row are log-linear plots, whereas the lower row are log-log plots.
    The color gradient for each value of $\rho$ indicates the change of the bulk mass $m^2\ell^2$: darker tones correspond to smaller values of the bulk mass $m^2\ell^2$.
    The bulk masses range from $0.00$ to $1.00$ in increments of $0.20$ from top to bottom.
    Dashed lines indicate fits according to Eq.~\eqref{eq:GabDgraph} in the upper row, and according to Eq.~\eqref{eq:GabDbound} in the lower row.}
    \label{fig:Percentages}
\end{figure*}

A generic correlation function  on the flake $\mathcal{G}$, such as that for the field $\phi_\mu$ described by the short-ranged action defined in Eq.~(\ref{mod1}), decays exponentially with the graph distance according to
\begin{align}
 \label{holo6} G_{\mu \nu} \sim e^{-D_{\mu\nu}/\xi}.
\end{align}
Herein, the correlation length $\xi$ depends on the system parameters such as the underlying lattices or the mass parameter $\hat{m}^2$ of the action. 
The exponential behavior in Eq.~(\ref{holo6}) is expected for both Euclidean and hyperbolic graphs. 
However, when expressed in terms of the boundary distance $d_{ab}$, we get quite distinct correlation functions.
In the Euclidean case, correlations remain exponentially suppressed due to the linear relation between the distance measures in Eq.~\eqref{holo4},
\begin{align}
 \label{holo7} G_{ab} \sim e^{-d_{ab}/\xi}\ \text{(Euclidean)},
\end{align} 
while in the hyperbolic case, the logarithmic scaling yields a power-law decay given by
\begin{align}
 \label{holo8} G_{ab} \sim \frac{1}{(d_{ab})^{2\Delta}}\ \text{(hyperbolic)}.
\end{align}
The power-law behavior in the hyperbolic case is reminiscent of the behavior of two-point correlation functions at second-order phase transitions, or more generally CFTs.
The scaling exponent $\Delta$ in our rough estimate derives from $\Delta \propto \xi^{-1}$, hence is determined by the systems parameters and, as such, is tunable within limits.

The discrete holographic correspondence laid out in Ref.~\cite{Dey2024} relates the bulk dynamics of $\phi_\mu$ to CFT-like correlation functions of some field or operator $\mathcal{O}_a$ on the boundary. Importantly, the fields $\mathcal{O}_a$ are not given by a simple limit of the bulk fields $\phi_\mu$, but rather appear as abstract fields on a one-dimensional space that is equivalent to $\partial\mathcal{G}$. For the sake of intuition, however, it is often useful to imagine that the operators $\mathcal{O}_a$ do actually live on the boundary $\partial\mathcal{G}$.

The associated connected two-point correlation functions of $\mathcal{O}_a$ are
\begin{equation}
    \langle \mathcal{O}_a \mathcal{O}_b \rangle = -M_{ab} + \sum_{i,j} M_{ai} G_{ij} M_{jb},
    \label{eq:TwoPointCorrFunct}
\end{equation}
where the bare inverse propagator is defined as  ${M_{\mu\nu} = \left(G^{-1}\right)_{\mu\nu} = -A_{\mu\nu} + \hat{m}^2 \delta_{\mu\nu}}$ and $G_{ij}$ is the bulk propagator~\cite{Dey2024}.
For two distinct boundary sites $a \neq b$, the inverse bare propagator reduces to the boundary component of the adjacency matrix $M_{ab} =  -A_{ab}$, because the mass term $\propto \delta_{ab}$ vanishes, leading to
\begin{equation}
\langle \mathcal{O}_a \mathcal{O}_b \rangle = A_{ab} + \sum_{i,j} A_{ai} G_{ij} A_{jb}\,.
\label{eq:TwoPointCorrS}
\end{equation}
Furthermore, given the two boundary sites $a$ and $b$, there are only two bulk sites $i_0(a)$ and $i_0(b)$ that are connected to them, i.e., $A_{ai_0(a)}=1$ and $A_{bi_0(b)}=1$, as illustrated in Fig.~\ref{fig:BulkBoundary}c).
Therefore, we have
\begin{equation}
  \langle \mathcal{O}_a \mathcal{O}_b \rangle = A_{ab} + G_{i_0(a)i_0(b)},
\end{equation}
where the propagator $G_{i_0(a)i_0(b)}$ is only evaluated on bulk sites.

The boundary two-point correlation functions in the hyperbolic-to-Euclidean crossover $\{7,3\} \to\{6,3\}$ are shown in Fig.~\ref{fig:Percentages} as a function of the graph distance $D_{ab}$ (first row) and of the boundary distance $d_{ab}$ (second row), using the same flakes as studied for the relation between distance measures.
An approximation for the boundary two-point correlation function is the propagator at the boundary $ \langle \mathcal{O}_a \mathcal{O}_b \rangle\simeq G_{ab}$~\cite{Dey2024},  
which, in the hyperbolic limit, exhibits exponential decay as a function of the graph distance according to
\begin{align}
  \langle \mathcal{O}_a \mathcal{O}_b \rangle \sim e^{-D_{ab}/\xi},
  \label{eq:GabDgraph}
\end{align}
and power-law decay as a function of the boundary distance according to
\begin{align}
\langle \mathcal{O}_a \mathcal{O}_b \rangle \sim \frac{1}{(d_{ab})^{2\Delta}}\,.\label{eq:GabDbound}
\end{align}
In Fig.~\ref{fig:Percentages}, the purely hyperbolic case, $\rho=0$, follows the expected scaling given in Eqs.~\eqref{eq:GabDgraph} and \eqref{eq:GabDbound}.
As the fraction of hexagonal defects increases, small deviations appear but the overall behavior remains consistent with the hyperbolic fit even at large $\rho$.
Only at large distance measures, deviations from the hyperbolic scaling become more noticeable in both cases.
Thus, Euclidean defects can be incorporated as a convenient approximation when constructing hyperbolic lattices, reducing the total number of sites without significantly altering boundary two-point correlations, consistent with Fig.~\ref{fig:Distances}.
In Appendix~\ref{app:Accuracy}, we further quantify the relation between the accuracy of the holographic scaling behavior and the reduction in the number of lattice sites induced by Euclidean defects.

\begin{figure}
    \centering
    \includegraphics[width=\linewidth]{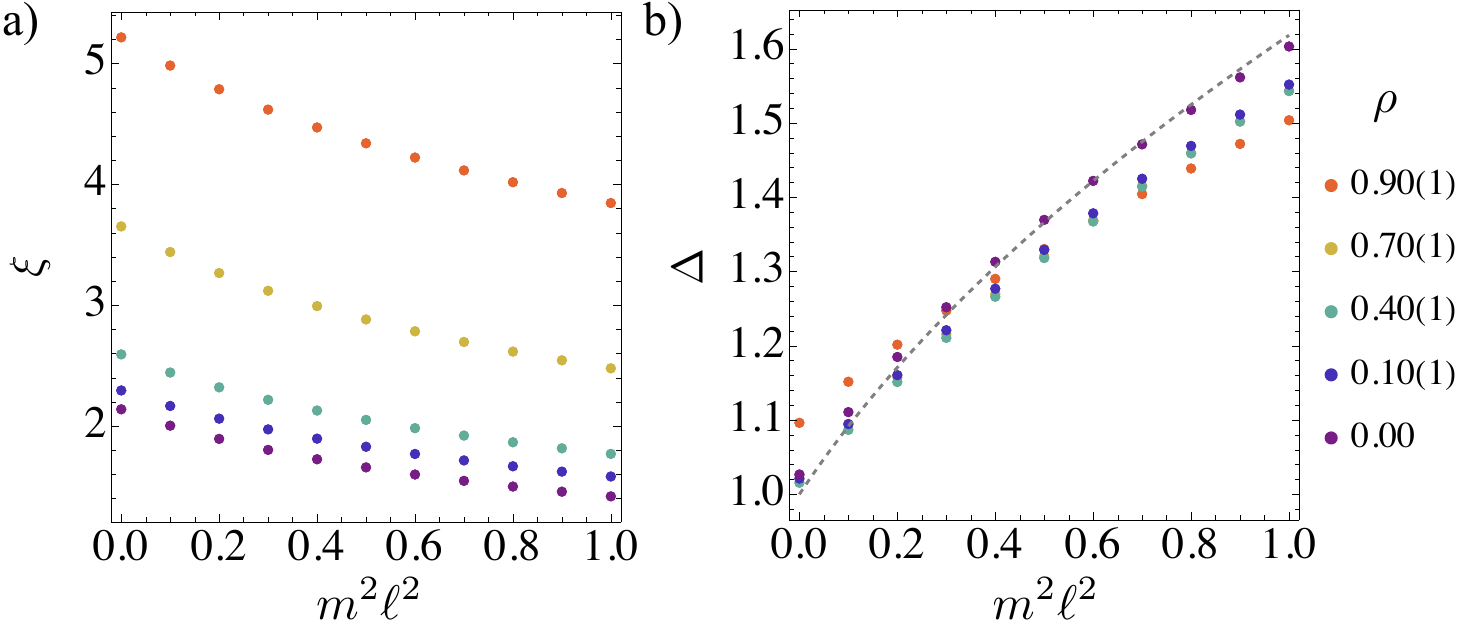}
    \caption{Correlation length $\xi$ and scaling dimension $\Delta$ as a function of the bulk mass $m^2\ell^2$ [with $\hat{m}^2=q+qh^2m^2\ell^2$, see Eq. (\ref{masseq1})]. The quantities shown are obtained from fitting $\langle \mathcal{O}_a \mathcal{O}_b \rangle$ computed via Eq. (\ref{eq:TwoPointCorrFunct}) to Eqs.~\eqref{eq:GabDgraph} and \eqref{eq:GabDbound}, respectively, in the non-Euclidean regime of the crossover with $\rho<1$. Different values of  $\rho$ are represented by different colors.
    The gray dashed line in the right panel is the curve $m^2 \ell^2=\Delta(\Delta-1)$, which accurately describes the data for hyperbolic lattices (purple points) and gives a useful estimate for defective lattices in the crossover regime.}
    \label{fig:XiDeltaMass}
\end{figure}

From Fig.~\ref{fig:Percentages}, we extract the correlation length $\xi$ and the scaling dimension $\Delta$ from fitting Eqs.~\eqref{eq:GabDgraph} and \eqref{eq:GabDbound}.
Their dependence on the bulk mass~$\sim m^2\ell^2$ is shown in Fig.~\ref{fig:XiDeltaMass}.
In the hyperbolic regime, the results obey the holographic relation between the scaling dimension $\Delta$ and the bulk mass $m^2\ell^2$ of the form ${m^2\ell^2=f[\Delta(\Delta-1)]}$, which in general deviates from the continuum limit $f_\mathrm{cont}(X)=X$ due to discretization effects \cite{Stegmaier2022,Brower2022,Dey2024}.
For coordination number $q=3$, however, the hyperbolic relation coincides with the one for the continuum limit, i.e., ${m^2 \ell^2=\Delta(\Delta-1) + \mathcal{O}(h)}$.
Fig.~\ref{fig:XiDeltaMass}b) shows $\Delta$ as a function of $m^2\ell^2$ for the hyperbolic-to-Euclidean crossover $\{7,3\}\to\{6,3\}$, where the pure hyperbolic regime gives a critical power-law scaling.

In our analysis, we have averaged the correlation functions in the crossover regime $0< \rho < 1$ over 30 realizations of flakes to highlight the effects expected on average. 
For individual realizations of flakes with a fraction $\rho$ of hexagons, as employed in experiments, this averaging does not apply. 
We find that the qualitative features and robustness of holographic boundary correlations against introducing defective tiles remains valid even for individual realizations, although deviations from the hyperbolic limit are more pronounced.
As an illustrative example, in Appendix~\ref{app:Alternated} we consider a specific defect distribution in which every ring contains both heptagons and hexagons in an alternating fashion, resulting in $\rho=0.5$. 
We show that the resulting boundary two-point correlation functions remain consistent with the disorder-averaged behavior obtained for lattices with defect fraction $\rho=0.5$.
It is possible that certain individual realizations minimize these deviations, for instance, by arranging the defects in a specific fashion, and may thus be particularly suited for experiments. 
We leave the exploration of this possibility for future work.

\subsection{Density of states}

The density of states (DOS) is a key spectral quantity that characterizes the number of available quantum states per energy level.
Since the DOS is a single-particle property, it is determined entirely by the quadratic terms of the Hamiltonian or action and, consequently, higher-order interaction terms do not contribute to the DOS.
The quadratic part of the action in Eq.~(\ref{mod1}) can be diagonalized in the eigenbasis of the adjacency matrix $A$ of the underlying graph $\mathcal{G}$. 
Defining the single particle Hamiltonian as an $N\times N$ matrix $H=-A$, the energy spectrum is then obtained by solving the eigenvalue problem
\begin{align}
 \label{dos1} \sum_{\nu} H_{\mu \nu}\psi^{(\alpha)}_\nu = \vare_\alpha \psi^{(\alpha)}_\mu,
\end{align}
where $\psi_\mu^{(\alpha)}$ and $\varepsilon_\alpha$ are the eigenfunctions and eigenvalues of the Hamiltonian, respectively, with $\alpha=1,\dots,N$ denoting the eigenstates and $\mu, \nu$ the site indices.
In the basis labeled by $\alpha$, the action becomes
\begin{align}
 \label{dos2} S = \frac{1}{2}\sum_{\alpha=1}^N (\vare_\alpha+\hat{m}^2)\phi_\alpha\phi_\alpha+S_{\rm int},
\end{align} 
such that the energy levels (eigenvalues) are shifted by the mass $\hat{m}$.
This shift only redefines the reference point of the spectrum, but does not modify its shape.

On periodic Euclidean tilings, the label $\alpha$ comprises the crystal momentum and band index, but on a general graph or flake $\mathcal{G}$ such a simple interpretation does not exist. 
However, a few general statements can be made about the spectrum on the flakes considered here. Since the coordination number of each site is either $q=3$ or less, the spectrum $\{\vare_\alpha\}$ is contained in the interval $[-3,3]$. 
For bipartite graphs, such as pure $\{6,3\}$- and $\{8,3\}$-flakes, or graphs in the crossover $\{8,3\}\to\{6,3\}$, the spectrum is symmetric in the sense that if $\vare_\alpha$ is an eigenvalue, then so is $-\vare_\alpha$. For non-bipartite graphs such as pure $\{7,3\}$-flakes or graphs in the crossover $\{7,3\}\to \{6,3\}$, this is not true.

The distribution of the eigenvalues $\{\varepsilon_\alpha\}$ defines the energy spectrum of the system, from which one can construct the density of states as
\begin{align}
 \label{dos3} D(E) \propto \sum_{\alpha=1}^N \bar{\delta}(E-\vare_\alpha),
\end{align}
where $\bar{\delta}(E)$ is a sufficiently regularized $\delta$-function. We choose $\bar{\delta}(E)=1$ if $-\frac{\delta E}{2}\leq E<\frac{\delta E}{2}$ and $\bar{\delta}(E)=0$ otherwise, with $\delta E=0.1$, and confirm that reasonable variations in the width $\delta E$ do not alter the result qualitatively. The DOS is normalized such that $\int \mbox{d}E\ D(E)=1$. For the flakes considered here, the DOS is nonzero only in the interval $E\in[-3,3]$ and the function is symmetric, $D(E)=D(-E)$, if and only if the graph is bipartite.

\begin{figure}[t]
    \centering
    \includegraphics[width=8.6cm]{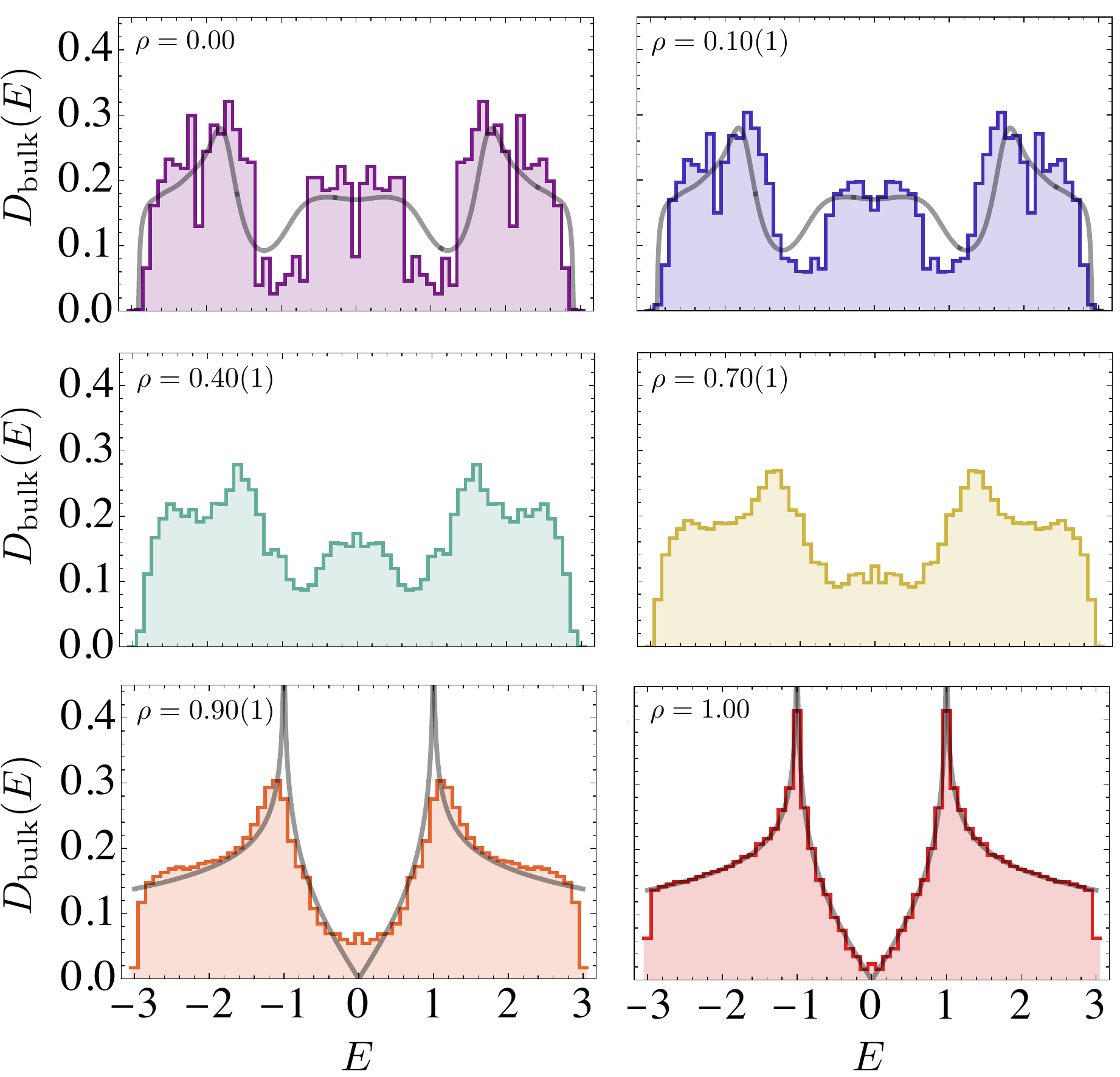}
    \caption{Bulk DOS in the hyperbolic-to-Euclidean crossover $\{8,3\}\to \{6,3\}$, where $\rho$ is the average number of hexagons. The DOSs for $\rho=0,1$ have been computed with flakes with $N\sim 10^4$ sites, the other plots with $0<\rho<1$ are each determined for 30 realizations with $N\sim 5\times 10^3$ sites. We observe that the DOS is regular within the crossover and similar in shape to the hyperbolic limit. The distinct features of the honeycomb lattice DOS are only developed as $\rho \sim 0.95$. 
    The limiting curves for pure $\{8,3\}$ and $\{6,3\}$ flakes are shown in gray. Deviations from the gray curve in the hyperbolic limit result from finite size effects in the bulk DOS, which, however, cannot be removed by studying larger flakes \cite{chen2023hyperbolic}.}
    \label{FigDOS83}
\end{figure}
\begin{figure}[t]
    \centering
    \includegraphics[width=8.6cm]{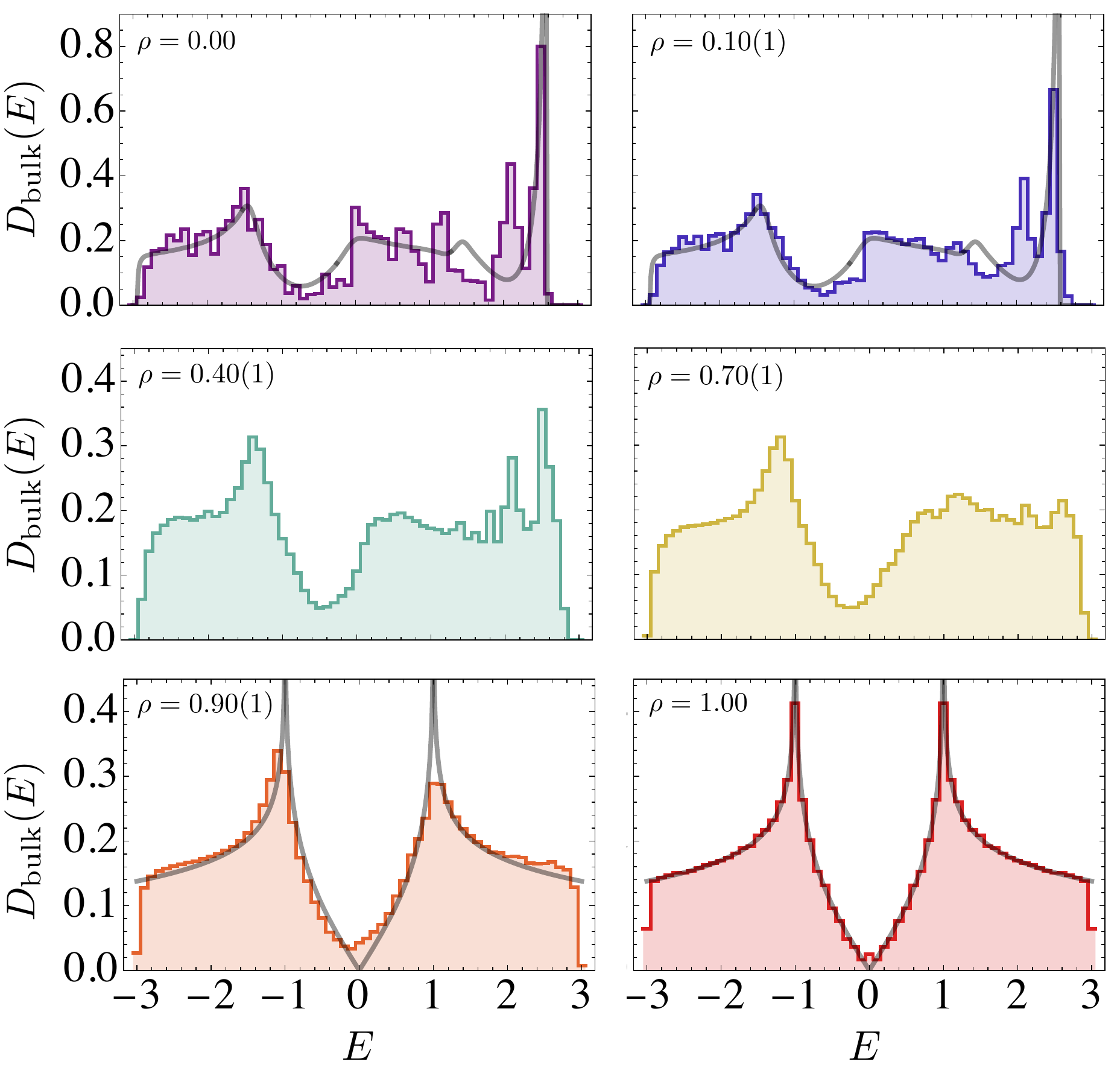}
    \caption{Bulk DOS in the hyperbolic-to-Euclidean crossover $\{7,3\}\to \{6,3\}$. Plots are determined for similar parameters as in Fig.~\ref{FigDOS83}, with the hyperbolic limit computed for ${N\sim 1.5 \times 10^4}$ sites. The distributions are asymmetric since the flakes within the crossover are not bipartite, unless $\rho=1$. The sharp peak of contributions around $E\simeq 3$ is eliminated for $\rho \sim 0.1$, while other features of the hyperbolic limit are more robust towards hexagon defects.
    The limiting curves for pure $\{7,3\}$ and $\{6,3\}$ flakes are shown in gray. Deviations from the gray curve in the hyperbolic limit result from finite size effects in the bulk DOS, which, however, cannot be removed by studying larger flakes \cite{chen2023hyperbolic}.}
    \label{FigDOS73}
\end{figure}

The DOS on an infinitely extended $\{p,q\}$ lattice is a mathematically well-defined object and can be computed for Euclidean and hyperbolic lattices. 
However, for hyperbolic lattices the result does not agree with the ${N\to \infty}$ limit of Eq.~(\ref{dos3}) computed for finite flakes, since for any number of sites $N$, a macroscopic fraction of $\mathcal{O}(N)$ sites resides at the boundary. 
For instance, for pure $\{8,3\}$ and $\{7,3\}$ flakes, the portion of boundary sites is $0.73N$ and $0.62N$, respectively.
These boundary sites contribute to the DOS and make it deviate from the limit of a boundary-free, infinitely extended lattice. 
For Euclidean flakes, while boundary effects are subleading in $N$ and thus technically vanish for $N\to \infty$, spurious boundary effects can be visible in the DOS if $N$ is not large enough.
For instance, for $\{6,3\}$ flakes, boundary modes contribute a significant weight at $E=0$.
Furthermore, for hyperbolic $\{p,3\}$ lattices, while the second moment of the DOS for the infinite lattice agrees with the constant coordination number 3, one obtains a smaller value for $\{p,3\}$-flakes of any finite size, because the average coordination number of sites is between 2 and 3. 
This behavior persists for the higher moments, and thus the full DOS, explaining the deviations between bulk- and flake-DOS for hyperbolic lattices.

To eliminate the contribution of boundary modes for finite-size flakes, the local or bulk DOS has proven useful~\cite{chen2023hyperbolic}. It is defined as
\begin{align}
 \label{dos4} D_{\rm bulk}(E) \propto \sum_{\alpha=1}^N \sum_{\mu \in \text{B}} \bar{\delta}(E-\vare_\alpha) |\psi^{(\alpha)}_\mu|^2,
\end{align}
where $\text{B}\subset \mathcal{G}$ is a subset of the full graph $\mathcal{G}$, representing the bulk.
For the analysis of the DOS here, we define $\text{B}$ as all sites of the graph with coordination number 3. 
Note that this is different from the definition of the bulk as $\mathring{\mathcal{G}}$ used before in Section~\ref{sec:Holography}.
Setting $\text{B}=\mathcal{G}$ recovers $D(E)$, while excluding some sites in $\text{B}$ effectively removes their weight from the DOS.
Since boundary effects introduce additional eigenstates absent in the bulk, the LDOS provides an approximate way to isolate bulk contributions.

In Figs. \ref{FigDOS83} and \ref{FigDOS73}, we show the hyperbolic-to-Euclidean crossovers in the bulk DOS for $\{8,3\}\to\{6,3\}$ and $\{7,3\}\to\{6,3\}$ as the average number of hexagons $\rho$ is changed from zero to one. 
The DOSs are computed for the specific flake parameters listed in Table~\ref{TabDOS83}.

In contrast to the boundary two-point correlation functions, the bulk DOS is more sensitive to the precise spatial arrangement of defects. 
As shown in Appendix~\ref{app:Alternated}, non-random flakes with alternating polygons along the rings and $\rho=0.5$ exhibit visible deviations from the disorder-averaged DOS, despite displaying similar boundary correlation behavior. 
This indicates that spectral properties are more affected by local geometric details than the holographic boundary correlations.

Before discussing the crossover regimes, we focus on the pure Euclidean and hyperbolic limiting cases. 
For the DOS of the infinite honeycomb lattice \cite{Kogan2021-fy}, define 
\begin{align}
 Z_0(E) &= \begin{cases}(3-|E|)(|E|+1)^3 & |E|<1, \\ 16|E| & 1\leq |E|\leq 3, \end{cases}\\
 Z_1(E) &= \begin{cases} 16|E| & |E|<1, \\ (3-|E|)(|E|+1)^3 & 1\leq |E|\leq 3, \end{cases}
\end{align}
and denote by $K(k^2)$ the complete elliptic integral of the first kind given by
\begin{align}
 K(k^2) =  \int_0^1\frac{\mbox{d}t}{\sqrt{(1-t^2)(1-k^2t^2)}}\, .
\end{align}
Then,
\begin{align}
 D(E) &=\frac{2|E|}{\pi^2\sqrt{Z_0(E)}} K\Biggl(\frac{Z_1(E)}{Z_0(E)}\Biggr)
 \end{align}
for $-3\leq E\leq 3$. The DOSs for the infinite $\{7,3\}$ and $\{8,3\}$ lattices were determined to high accuracy with the continued fraction method in Ref.~\cite{PhysRevB.108.035154}, which contains the explicit expressions. 
While the DOS of the $\{6,3\}$ lattice is recovered well by the bulk DOS with $N\sim 10^4$ sites, as shown in Figs.~\ref{FigDOS83} and \ref{FigDOS73}, the agreement for the hyperbolic lattices is only qualitative for similar values of $N$. 
We emphasize, however, that increasing $N$ in the hyperbolic case does not improve the result, which is in fact already converged. 
This showcases that the bulk DOS, while giving a workable estimate of the DOS on the infinite lattice, is only approximate and fails in cases of extreme numbers of boundary sites. 
Importantly, the bulk DOS removes the contribution of boundary states with $E=0$ in the case of the $\{6,3\}$ lattice, reproducing the exact value of $D(0)=0$ as $N\to \infty$.

While the bulk DOS cannot capture all features of the DOS in the hyperbolic limits, we expect it to be a good approximation of the infinite system for large enough $\rho$, since the curves look smooth and the graphs become more Euclidean-like with less sites on the boundary. In the case of the $\{8,3\}\to\{6,3\}$ crossover, where all intermediate graphs are bipartite, we observe that the crossover DOS is a regular function without singularities and gaps. Its shape approximately resembles that of the hyperbolic limit for sufficiently large $\rho \lesssim 0.5$, while the characteristic features of the honeycomb DOS, i.e., the van Hove singularities $|E|=1$ and the linear behavior $D(E)\propto |E|$ for $E=0$, are only developed for $\rho \gtrsim 0.95$. Similarly, in the $\{7,3\}\to\{6,3\}$ crossover, the intermediate DOS is regular and resembles features of the hyperbolic limit until $\rho \lesssim 0.5$. The sharp peak around $E\simeq 3$ of the $\{7,3\}$ lattice is eliminated for $\rho \sim 0.1$, while the asymmetric shape $D(E)\neq D(-E)$ is persistent up to $\rho \sim 0.9$.


\begin{table}[h!]
\begin{center}
\begin{tabular}{|c|c|c|c|}
\hline
 \multicolumn{4}{|c|}{$\{8,3\} \to \{6,3\}$}\\
\hline\hline
\ $\rho$ \ & \ $l$ \  & \ no. of realizations \  & \ $N$ \  \\
\hline
\ 0 \ & \ 6 \  & 1 & 10800  \\
\hline
 0.10(1) & 5 & 30 & 2300(200)  \\
\hline
 0.20(1) & 6 & 30 & 6300(700)  \\
\hline
 0.30(1) & 6 & 30 & 4500(900)  \\
\hline
 0.40(1) & 6 & 30 & 3400(600)  \\
\hline
 0.50(1) & 7 & 30 & 5900(1400)  \\
\hline
 0.61(1) & 7 & 30 & \ 4000(1000) \   \\
\hline
 0.70(1) & 8 & 30 & 4500(1400) \\
\hline
 0.80(1) & 9 & 30 & 4600(1400)  \\
\hline
\ 0.90(1) \ & \ 12 \  & 30 & 5600(1600)  \\
\hline
\ 1 \ & \ 41 \  & 1 & 10086  \\
\hline\hline
 \multicolumn{4}{|c|}{$\{7,3\} \to \{6,3\}$}\\
\hline\hline
\ $\rho$ \ & \ $l$ \  & \ no. of realizations \  & \ $N$ \  \\
\hline
\ 0 \ & \ 8 \  & 1 & 15435  \\
\hline
 0.10(1) & 7 & 30 & \ 4500(400) \  \\
\hline
 0.20(1) & 7 & 30 & 3600(400)  \\
\hline
 0.30(1) & 8 & 30 & 6500(700)  \\
\hline
 0.40(1) & 8 & 30 & 4800(600)  \\
\hline
 0.50(1) & 8 & 30 & 3400(600)  \\
\hline
 0.60(1) & 9 & 30 & 4400(700)   \\
\hline
 0.70(1) & \ 10 \ & 30 & 4800(700)  \\
\hline
 0.80(1) & 11 & 30 & 3900(700)  \\
\hline
\ 0.90(1) \ & \ 14 \  & 30 &  5000(900)  \\
\hline
\ 1 \ & \ 41 \  & 1 & 10086  \\
\hline
\end{tabular}
\end{center}
\caption{Parameters for the hyperbolic-to-Euclidean $\{8,3\} \to \{6,3\}$ and $\{7,3\} \to \{6,3\}$ crossover flakes. 
For each average fraction of Euclidean defects $\rho$, with standard deviation in parentheses, we build flakes with $l$ concentric rings. 
In the intermediate regime, 30 independent realizations are generated for each value of $\rho$. 
The total number of sites, $N$, is given together with the standard deviation in parentheses.}
\label{TabDOS83}
\end{table}

\section{Conclusions and Outlook}

We have studied the crossover from hyperbolic to Euclidean lattices by constructing hyperbolic flakes containing Euclidean defects in a controlled way via the tile-by-tile inflation method, which provides a versatile and systematic way to construct lattices with arbitrary combinations of polygons with a fixed coordination number.
This method allows us to explore a broad class of geo\-metries, including smooth transitions from hyperbolic to Euclidean lattices such as the $\{7,3\}\to\{6,3\}$ and $\{8,3\}\to\{6,3\}$ crossovers.
Analyzing the relation between boundary and graph distance measures, boundary two-point correlation functions, and the bulk density of states across different steps of the crossover provides insights into how local Euclidean deformations affect boundary and bulk observables in negatively curved lattices.

A central result of our study is the remarkable robustness of boundary observables under the introduction of Euclidean defects.
The analysis of boundary two-point correlation functions reveals that the introduction of a moderate amount of Euclidean defects in hyperbolic lattices does not significantly alter the boundary behavior.
This is supported by the validity of the exponential relation between the graph and boundary distance measures over a wide range of defect fractions.
Noticeable deviations from this hyperbolic scaling appear only at short distances or for large defect concentrations. As a next step, we could examine whether the influence of defects increases if the defect polygons have smaller $p$, i.e., positively curved defects like pentagons, or if the underlying hyperbolic lattice has larger $p$ while maintaining hexagons as the defect structures.

In contrast, the bulk density of states exhibits a more pronounced response to the increasing fraction of Euclidean defects.
For the bipartite $\{8,3\}\to\{6,3\}$ and the non-bipartite $\{7,3\}\to\{6,3\}$ crossovers, the intermediate DOS retains signatures of the defect-free hyperbolic lattices up to intermediate defect fractions of $\rho \lesssim 0.5$. 
In particular, in the $\{7,3\}\to\{6,3\}$ case, the overall asymmetry of the $\{7,3\}$ DOS remains robust even at large defect concentrations, whereas the sharp peak characteristic of the $\{7,3\}$ lattice is rapidly suppressed.
While this suppression is consistent with disorder-induced broadening and smoothening of DOS peaks due to breaking of translational symmetries, as observed for van Hove singularities~\cite{PhysRevB.66.193311,PhysRevResearch.3.L032067,PhysRevB.105.075144}, the true mechanism for sharp DOS features disappearing upon introducing defects here poses an important question for future work. Furthermore, distinctive Euclidean features such as van Hove singularities and the linear dispersion relation at zero energy emerge only when Euclidean defects dominate the lattice close to $\rho\sim0.9$.
In future work, one could study the DOS with periodic boundary conditions by connecting the boundary sites of the flakes, which would solve the problem of edge effects.
Overall, our results demonstrate that bulk properties are more sensitive to local geometric deformations, while boundary properties remain remarkably robust for moderate levels of Euclidean defects.

Beyond providing insight into curvature-driven spectral and correlation effects, our results have practical implications for numerical and experimental realizations of hyperbolic lattices. 
The exponential growth of number of sites with graph diameter is a  major obstacle in the experimental realization of purely hyperbolic lattices. For instance, vertex numbers in experiments using circuit QED and coplanar waveguides \cite{kollar2019hyperbolic,fleury2024,Xu25} have so far been limited to $\sim50-150$, some of the challenges for scaling being the need to accurately adjust local on-site mass terms $\hat{m}^2_\mu\sim\mathcal{O}(10\ \text{GHz})$ to not impede the much smaller hopping terms $t_{\mu\nu}\sim \mathcal{O}(100\ \text{MHz})$, or the spatial extent of the coupling capacitors implementing $t_{\mu\nu}$, although the methods developed in Refs.~\cite{kollar2019hyperbolic,fleury2024,Xu25} show great promise to scale to bigger systems in the future. Comparable system sizes have been accomplished with topological photonic lattices in Ref.~\cite{huang2024hyperbolic}. In topoelectrical circuits \cite{Lenggenhager2021,zhang2022observation}, several hundreds of vertices have been achieved experimentally. The dynamical response in this classical platform is efficiently emulated numerically using LTSPICE software, which can, in principle, scale to much bigger systems. They are, therefore, a suitable platform to experimentally or computationally study the semi-classical effects as a function of $\rho$ studied here in a realistic setting. For the purpose of realizing hyperbolic lattices, they would benefit less from using defects than other platforms, because of the favorable scaling.

Our findings show that essential boundary physics, such as those relevant to holographic dualities, can be faithfully reproduced in lattices with small vertex numbers and a controlled amount of Euclidean defects. 
This constitutes a promising way to emulate boundary observables of hyperbolic lattices without requiring to construct pure hyperbolic lattices, which are computationally and experimentally expensive to build.

\section*{Acknowledgments} 

The authors thank Anffany Chen, Michael M. Scherer, and Ronny Thomale for insightful comments. 
MTS acknowledges funding and hospitality from the Theoretical Physics Insitute (TPI) at the University of Alberta, and funding from the Deutsche Forschungsgemeinschaft (DFG, German Research Foundation) within Project-ID 277146847, SFB 1238 (project C02). 
IB acknowledges funding from the Natural Sciences and Engineering Research Council of Canada (NSERC) Discovery Grants RGPIN-2021-02534 and DGECR2021-00043.

\begin{appendix}

\section{Site counting on a $\{p,3\}$ hyperbolic lattice}\label{app:Recurrence}

The up-down sequences can be used to determine the number of sites on each ring of the $\{p,3\}$ lattice.
For this, denote the number of up-sites on the $n$th ring by $a_n$, and the number of down-sites by $b_{n}$. 
The procedure described above yields the recurrence relations 
\begin{align}
 \label{graph8} a_{n} &= (p-4) a_{n-1}-b_{n-1},\\
 \label{graph9} b_n &= a_{n-1},\\
 \label{graph10} a_1&=p,\ b_1 =0,
\end{align}
which can be written as $a_n = (p-4)a_{n-1}-a_{n-2}$ with initial conditions $a_1 =p$, $a_2 = p(p-4)$. The recursion for $p>6$ is solved by
\begin{equation}
\begin{split}
  a_n =& \frac{p}{\sqrt{(p-2)(p-6)}}\Biggl[\Biggl( \frac{p-4+\sqrt{(p-2)(p-6)}}{2}\Biggr)^n \\
 \label{graph13} &- \Biggl(\frac{p-4-\sqrt{(p-2)(p-6)}}{2}\Biggr)^n\Biggr],
  \end{split}
\end{equation}
\begin{equation}
\begin{split}
  b_n =& \frac{p}{\sqrt{(p-2)(p-6)}}\Biggl[\Biggl( \frac{p-4+\sqrt{(p-2)(p-6)}}{2}\Biggr)^{n-1} \\
 \label{graph13b} &- \Biggl(\frac{p-4-\sqrt{(p-2)(p-6)}}{2}\Biggr)^{n-1}\Biggr].
 \end{split}
\end{equation}
The number of sites on the $n$th ring is
\begin{align}
 \label{graph15} N_{\rm ring}(n) = a_n + b_n,
\end{align}
and the total number of sites for a graph with $l$ rings is
\begin{align}
 \label{graph16} N(l) ={}& \sum_{n=1}^l N_{\rm ring}(n).
\end{align}
For $p\to 6$ we find $a_n=6n$. The average coordination number for $\{p,3\}$ lattices is defined as
\begin{align}
 \bar{q} = \lim_{l \to \infty}  \frac{3(N(l-1)+b_l)+2a_{l}}{N(l)}.
\end{align}

\section{Flakes}\label{app:flakes}
    
The flakes generated via the tile-by-tile inflation method and analyzed via the DOS and boundary two-point correlation functions for the hyperbolic-to-Euclidean crossovers ${\{8,3\} \to \{6,3\}}$ and ${\{7,3\} \to \{6,3\}}$ are characterized by the parameters provided in Table~\ref{TabDOS83}.

\begin{figure}[t!]
    \centering
    \includegraphics[width=\linewidth]{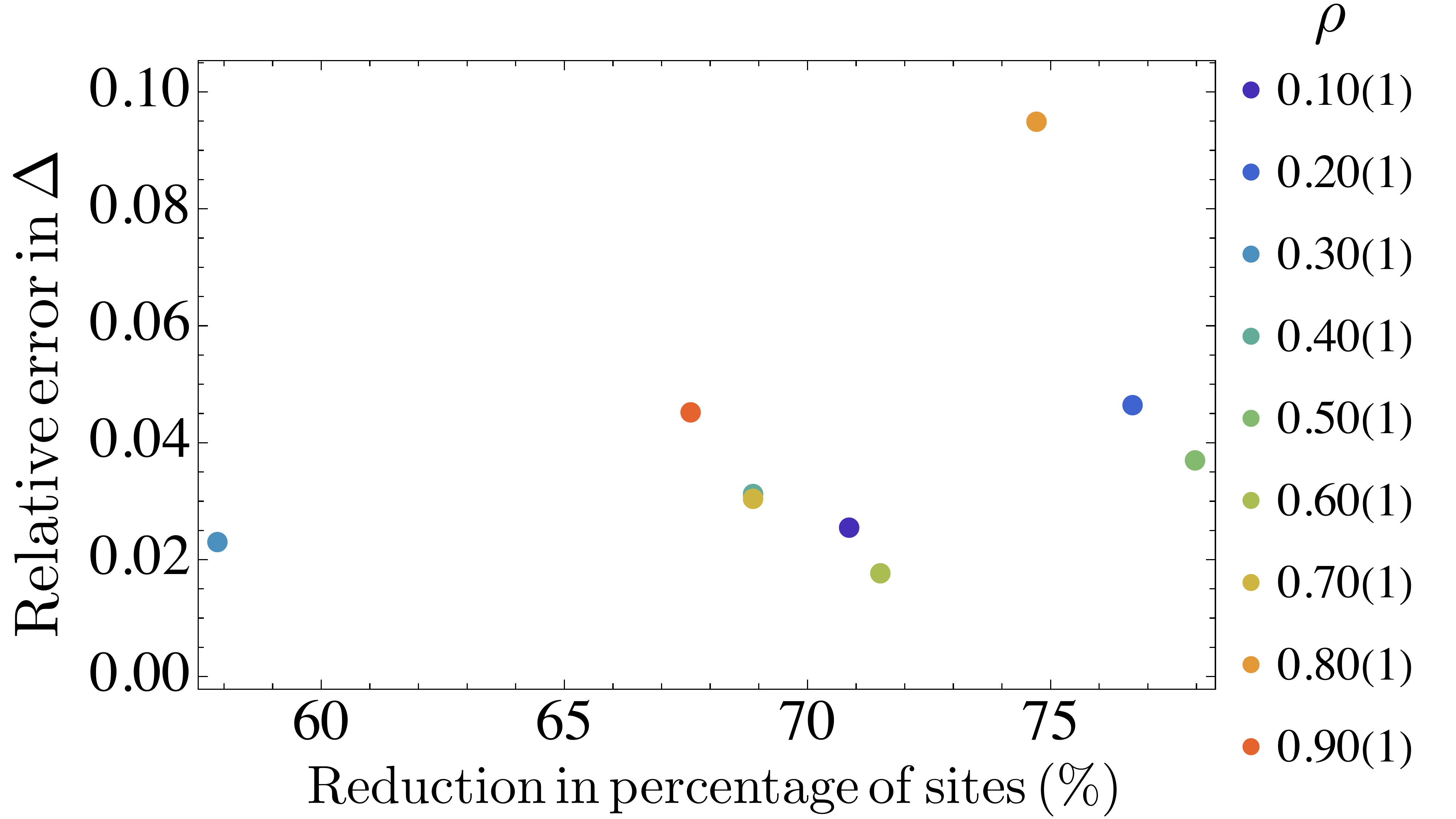}
    \caption{Relative error in the scaling dimension, ${|\Delta-\Delta_\mathrm{hyp}|/\Delta_\mathrm{hyp}}$, as a function of the relative reduction in the number of lattice sites, $1-N(\rho)/N_\mathrm{hyp}$.
    Different colors correspond to different defect concentrations $\rho$.
    Here, $\Delta_\mathrm{hyp}$ denotes the scaling dimension for the pure $\{7,3\}$ hyperbolic lattice with $l=8$ rings, corresponding to $N_\mathrm{hyp}=15435$ sites.
    The numbers of lattice sites for different values of $\rho$ are given in Table~\ref{TabDOS83}.
    The scaling dimension for different values of $\rho$ are extracted from Fig.~\ref{fig:XiDeltaMass}b.}
    \label{fig:RelativeError}
\end{figure}

\section{Scaling accuracy and site reduction}\label{app:Accuracy}

To quantify the effect of defects on the holographic scaling behavior, Fig.~\ref{fig:RelativeError} shows the relative error ${|\Delta-\Delta_\mathrm{hyp}|/\Delta_\mathrm{hyp}}$ as a function of the relative reduction in lattice size, $1-N(\rho)/N_\mathrm{hyp}$, with $N_{\rm hyp}=N(0)$. 
We observe that reductions of the number of lattice sites by roughly $60\%-80\%$ still lead to only small deviations in the extracted scaling dimension, typically at the few-percent level. 
This indicates that defective lattices can substantially reduce system size while maintaining the characteristic holographic scaling behavior.

\section{Comparison between random and alternating defect distributions}
\label{app:Alternated}

In the main part of our work, we consider random, uncorrelated insertion of hexagonal defects in the hyperbolic-to-Euclidean crossover and averaged the physical observables over 30 independent realizations for each value of the defect fraction $\rho$.
This averaging procedure samples a broad range of spatial configurations, including realizations in which defects are more clustered or more uniformly dispersed, thus providing statistically representative results. 
This naturally raises the question of whether the observed holographic boundary behavior and bulk observables persist for non-random or highly ordered defect distributions. 
To address this point, we study here two particular ordered realizations corresponding to alternating heptagon-hexagon and octagon-hexagon lattices at defect fraction $\rho=0.5$. For this, along each ring, hexagons alternate with the corresponding heptagons or octagons.

\begin{figure}
    \centering
    \includegraphics[width=8cm]{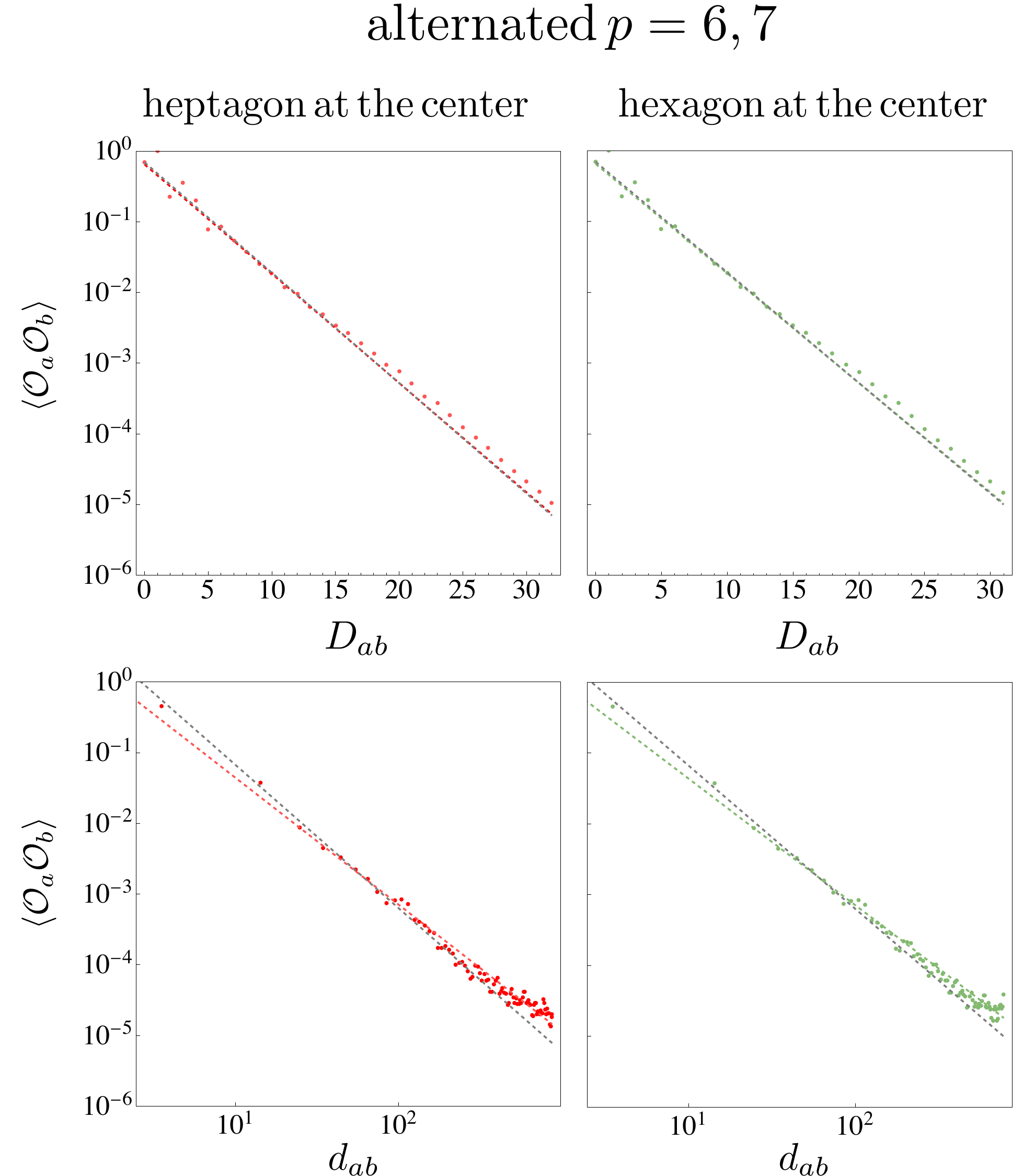}
    \caption{Boundary two-point correlation functions $\langle\mathcal{O}_a\mathcal{O}_b\rangle$ for alternating heptagon–hexagon lattices in the hyperbolic-to-Euclidean crossover at $\rho=0.5$.
    The upper row shows the correlation functions as a function of the graph distance $D_{ab}$, while the lower row shows them as a function of the boundary distance $d_{ab}$.
    The left panels correspond to realizations with a heptagon at the center (red), while the right panels correspond to realizations with a hexagon at the center (green). Colored dashed lines indicate fits according to Eqs.~\eqref{eq:GabDgraph} and~\eqref{eq:GabDbound} for the upper and lower rows, respectively. 
    The gray dashed lines show the corresponding fits obtained from averaging over the 30 random realizations considered in Fig.~\ref{fig:Percentages}.}
    \label{fig:2pointAlternating}
\end{figure}

\begin{figure}
    \centering
    \includegraphics[width=8.6cm]{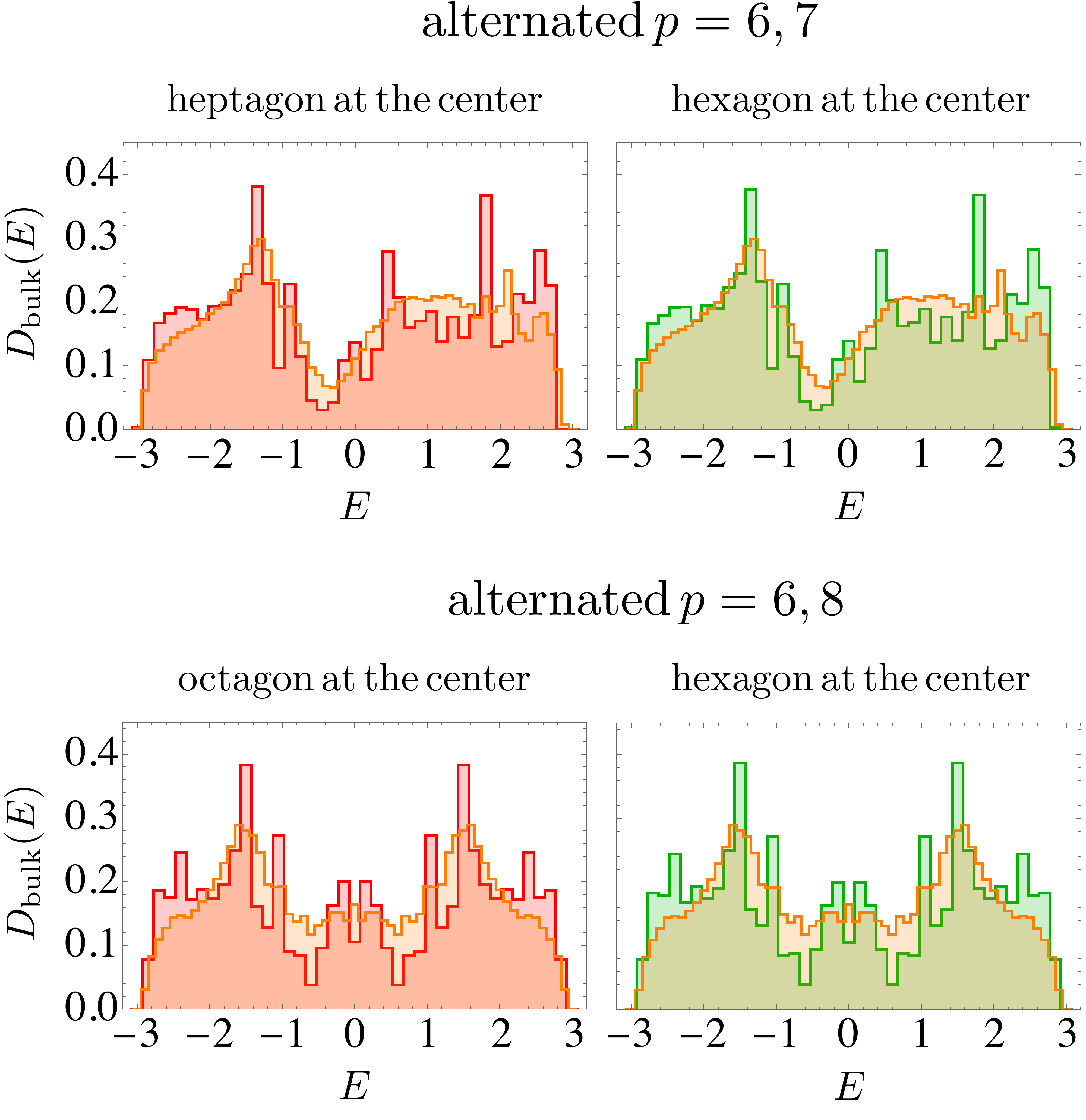}
    \caption{Bulk DOS in the hyperbolic-to-Euclidean crossover $\{7,3\}\to\{6,3\}$ (top panels) and $\{8,3\}\to\{6,3\}$ (bottom panels) at $\rho=0.5$.
    The orange curves show the disorder-averaged DOS obtained from the 30 random realizations considered in Figs.~\ref{FigDOS73} and~\ref{FigDOS83}, respectively.
    These results are compared with the corresponding alternating heptagon-hexagon (top panels) and octagon-hexagon (bottom panels) lattices. 
    The left panels correspond to realizations with a heptagon/octagon at the center (red), while the right panels correspond to realizations with a hexagon at the center (green).}
    \label{fig:DOSalternating}
\end{figure}

Figure \ref{fig:2pointAlternating} shows that the alternating lattices remain consistent with the disorder-averaged behavior for the boundary two-point correlation functions. 
In particular, the qualitative scaling behavior and the extracted fits remain largely unchanged despite the highly ordered nature of the defect arrangement. 
This suggests that the holographic boundary correlations are primarily controlled by the large-scale geometric crossover between hyperbolic and Euclidean regions, rather than by microscopic details of the defect distribution.
In contrast, the bulk DOS exhibits a stronger sensitivity to the precise defect arrangement. 
Figure \ref{fig:DOSalternating} compares the DOS obtained from alternating heptagon-hexagon and octagon-hexagon lattices with the corresponding disorder-averaged DOS at $\rho=0.5$. 
Noticeable quantitative deviations appear between the ordered and disorder-averaged realizations.
This distinction indicates that spectral observables are more strongly influenced by local geometric structure, whereas the boundary correlation functions display a substantially more robust behavior. 
Therefore, alternating lattices remain effective probes for detecting holographic boundary correlations even though their spectral properties differ from those of disorder-averaged configurations.

Finally, we observe that the choice of the central polygon being a 7-/8-gon versus a hexagon produces only minor differences. 
In both cases, the bulk DOS and the boundary two-point correlation functions remain nearly identical. 
These results support the robustness of the holographic boundary signatures against variations in the spatial distribution of defects.

\end{appendix}

\vfill
\newpage

\bibliography{refs_defects}

\end{document}